\documentclass{article}

\usepackage{arxiv}

\usepackage{standalone}
\usepackage{enumitem}
\usepackage{amsmath, amssymb, bm, graphicx, xcolor, hyperref, mathrsfs, stackrel}
\usepackage{xr}
\usepackage{xr-hyper}
\usepackage{comment}
\usepackage{authblk} 

\global\def\1{\boldsymbol{1}}

\title{Emergence of agriculture in an artificial society of reinforcement learning agents}

\author[a]{Gautier Hamon}
\author[b,c]{Mart\'i S\'anchez-Fibla}
\author[a, d]{Cl\'ement Moulin-Frier\thanks{Corresponding author: clement.moulin-frier@inria.fr}}
\author[e,f,g]{Ricard Sol\'e\thanks{Corresponding author: ricard.sole@upf.edu}}

\affil[a]{Flowers AI and CogSci lab, Inria, Université de Bordeaux, France}
\affil[b]{Department of Information and Communications Technologies, Universitat Pompeu Fabra, 08018, Barcelona, Spain}
\affil[c]{Artificial Intelligence Research Institute, IIIA (CSIC), Campus de la UAB, Bellaterra, Barcelona, 08193, Spain}
\affil[d]{BioTiC team (Inria, INSA Lyon, CITI, UR3720), France}
\affil[e]{Complex Systems Lab, Universitat Pompeu Fabra, Dr Aiguader 88, 08003 Barcelona, Spain}
\affil[f]{Instituci\'o Catalana de Recerca i Estudis Avançats, Llu\'is Companys 23, 08010 Barcelona, Spain}
\affil[g]{Santa Fe Institute, 1399 Hyde Park Road, Santa Fe NM 87501, USA}

\begin{document}
\maketitle

\begin{abstract}
The origin of agriculture represents a major evolutionary transition and a
paradigmatic example of how complex collective behaviors emerge from simple
interactions. Here we introduce an artificial society of reinforcement learning
agents embedded in a dynamic ecological environment to identify general
principles underlying this transition. Within this system, agricultural
practices emerge spontaneously—without explicit instruction—through the coupled
dynamics of learning and environmental modification. We show that this
transition is governed by four key ingredients: individual planning through the
valuation of delayed rewards, social vulnerability to cheaters, stabilization
via social learning, and an emergent lock-in effect that renders agriculture
effectively irreversible once established. In particular, we demonstrate that
social learning acts as a ``firewall'' that suppresses cheater invasion and
enables the propagation of successful strategies, leading to sustained
population growth and nonlinear amplification of domesticated resources.
Together, these results reveal universal mechanisms linking individual
decision-making, social interactions, and ecological feedbacks. More broadly,
they highlight the potential of artificial societies as experimental platforms
to study the emergence of cultural innovations and major evolutionary
transitions.
\end{abstract}


The rise of agriculture marks an evolutionary milestone for many species in terrestrial ecosystems, from insects \cite{mueller2005evolution} to humans \cite{rindos2013origins}. It is a striking example of ecosystem engineering in which a species transforms its environment in ways that profoundly impact the flows of matter and energy \cite{samson1996organisms,jones1997positive}. Ecosystem engineering can occur in various ways: organisms may influence their environment simply by existing within it, or they may actively reshape it to create more favorable conditions. Agriculture belongs to the latter, as it involves the deliberate cultivation of food sources to provide a reliable source of nourishment.

Agriculture has emerged independently multiple times across various taxa and at various scales. Through niche construction, humans, ants, termites, and beetles have successfully domesticated crops, livestock, or fungi \cite{purugganan2022domestication}. Comparative studies reveal a broad spectrum of agricultural strategies, from microbial pest management to artificial selection \cite{schultz2008major,nygaard2016reciprocal}. Despite this diversity, convergent evolution suggests the presence of shared constraints or opportunities \cite{schultz2008major,mcghee2022convergent}. These convergences are reflected in several key traits common to all groups: sowing, spatial niche engineering, crop protection through weeding, and the establishment of reciprocal mutualism \cite{purugganan2022domestication}. This mutualism is especially significant, as it defines a feedback loop where the domesticator regulates crop fitness and benefits from the resources it produces.

All systems displaying agriculture share key generative rules. They involve the intentional cultivation of specific resources, where organisms or environments are selected and modified to optimize growth and ensure continued yield. They operate through cycles of propagation, where resources are regularly sown, bred, or cultivated, followed by harvest, during which useful output is extracted. The system gains direct or indirect benefit from this process and adapts over time through feedback mechanisms that refine and improve the cultivation strategies. One consequence for groups of individuals would be the partial abandonment of the foraging strategy and a reduction in mobility. 

Domestication emerged independently in different places, suggesting that this transition is likely to occur once the right conditions are met. How likely is it? What kind of universal properties are expected to be shared? What are the roles of individual exploration versus collective patterns, as well as cognitive and environmental constraints? How does this collaborative phenomenon deal with cheaters? Once the domestication state is reached, is it locked-in?  Answering these questions would greatly benefit from a modeling approach in which a population of artificial learning agents, interacting with a given environment, can discover the crucial components that make domestication possible. 

In this paper, we show that key components of plant domestication can emerge in populations of reinforcement learning (RL) agents. Under favorable conditions, agriculture arises spontaneously in small groups that discover how to eco-engineer their environment by promoting the growth of cultivable plants, progressively abandoning foraging, and adopting sedentary behavior. However, in larger populations, this strategy becomes unstable due to the invasion of exploitative agents (“cheaters”), which undermine collective investment in cultivation. This destabilizing effect can be suppressed by introducing mechanisms that allow the propagation of domestication across generations, indicating that social learning acts as a firewall against cheaters. Under these conditions, the agricultural state becomes stable and effectively irreversible, consistent with a major ecological and cultural transition.

\begin{figure*}[t]
\begin{center}
    \includegraphics[width=0.95\linewidth]{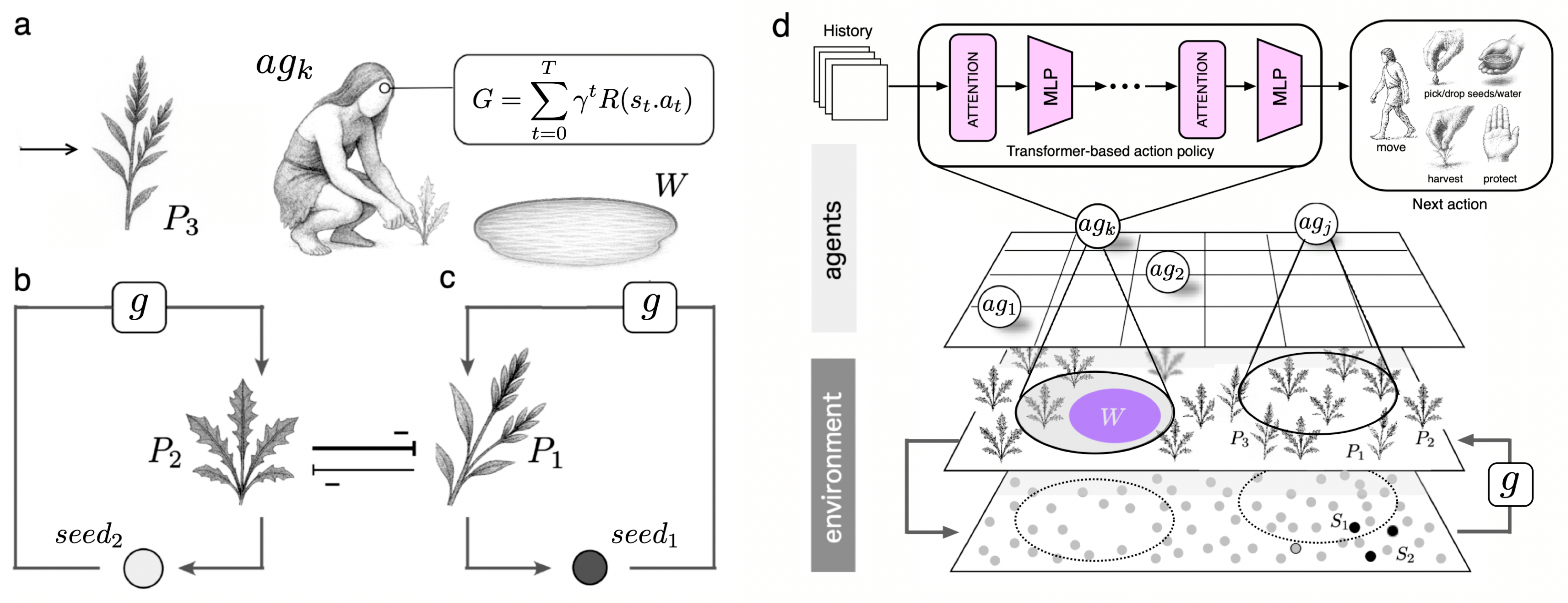}
\end{center}
    \caption{{\bf Modelling agent-environment interactions}. (a) Each agent $ag_k$ ($k=1,2,...,N$) starts with a random behavior interacting within a spatial landscape with three different kinds of plants ($P_i$) and water ($W$). One of them ($P_1$) is the most reward plant (e.g. because it provides more nutrients to the agent), but is not common and is ecologically overcompeted (b-c) by another plant ($P_2$), which provides no reward. They produce seeds ($ seed_1$, $ seed_2$) that need to germinate ($g$) and can benefit from external resources, such as water ($W$). Another plant ($P_3$) grows spontaneously, making it an easy resource to forage but providing low reward. Agents can navigate the map, observe their local neighbourhood and act in multiple ways on different plants, their seeds and water (picking and dropping those items over the map, as well as harvesting or consuming plants, (d)). When consuming a plant of a given species, an agent receives a scalar reward ($R$) associated with that species. The agent's objective is to discover a policy—a strategy for choosing actions based on observed states—that maximizes its individual cumulative reward over time, using Reinforcement Leaning (RL). The cumulative reward ($G$, closed box) depends on the sequences of both environmental states and actions, and is weighted by the parameter $\gamma$ (the discount factor) which reflects the agent’s temporal horizon or foresight. This \emph{Multi-Agent Reinforcement Learning (MARL)} system and its environmental context are summarised in (d), where agents navigate and act on the environment grid. The internal cognitive architecture of each agent is sketched on the top-right diagram (Transformer-based action policy).}
    \label{fig:setup}
\end{figure*}

\section{Methods}
\label{sec:methods}

Our modelling approach relies on a recent computational paradigm to study cognitive agents in social-ecological systems~\cite{Perolat2017, sanchez2024cooperative,hertz2025beyond,lu2024jaxlife}, relying on two main components: a simulated spatial environment with complex ecological dynamics and a population of agents adapting their behavior through multi-agent reinforcement learning. We describe it below and provide additional formalization detail in \textit{Material and Methods} (hereafter MM) and in Supplementary Material (hereafter SM).

\paragraph{Environment dynamics.} As sketched in Fig.~\ref{fig:setup}, a population of agents $\mathcal{P}=\{ag^1, \ldots ag^N\}$ interact with an environment $\cal E$ forming a discrete 2D spatial landscape. The environment is described in terms of the spatiotemporal dynamics of three types of plants, namely: (1) $P_1$ an uncommon but highly rewarding plant, (2) a common weed $P_2$ that will ecologically outcompete $P_1$ but does not provide a reward, and (3) $P_3$, a plant that enters the system randomly and provides low reward when consumed. In addition, a fixed patch of water ($W$ in Fig.~\ref{fig:setup}) is also introduced. Water acts as a fertilizer: When present in an environment cell $(x,y)$, it improves plant growth within its  $3 \times 3$ neighborhood. In soil, water evaporates with a given probability of disappearance per time step. $P1$ and $P2$ plants follow seasonal cycles: growing during summer, spreading seeds and dying at the start of winter.

At the beginning of summer, the seeds germinate with probability $\eta_k$ (see MM for definition). Seeds that do not sprout remain on the grid but can disappear with a given probability before the next season. During the last ten time steps of summer, plants $P_1$ and $P_2$ probabilistically disperse seeds in their $3\times3$ neighborhood, with hyperparameters controlling the spreading of each plant (see MM for detail). Multiple seeds of different plant species can occupy the same grid cell. 

At the beginning of winter, all plants die, leaving only seeds. At the beginning of the next summer, the seed that germinates on a given cell is categorically sampled (either no seed germinates, or only a $P_1$ or a $P_2$ seed). The hyperparameters of the underlying probability distribution control the germination competition between $P_1$ and $P_2$ and are chosen to provide an advantage to $P_2$ (see MM for details). During summer, $P_1$ nutritional value increases sigmoidally with time and with the amount of water received by the plant since the start of the season.
Formalization detail are provided in MM and full details of the environmental dynamics and all fixed environment parameter values in Section I of the SM, including grid initialization, water mechanics, and seed decay.


The simulated environment contains two edible plant species, $P_1$ and $P_3$.  $P_1$ provides the highest nutritional benefit (i.e. highest reward) but requires active management to limit competition with $P_2$ and prevent over-harvesting, which could lead to depletion. This corresponds to a \textit{Common Pool Resource} (CPR) scenario, where consumption by one agent reduces availability for others but access remains non-excludable~\cite{Perolat2017}. In contrast, $P_3$ represents a spatially random and passively available food source, providing a baseline resource without requiring active cultivation. This asymmetry in value and renewal mechanisms captures key features of real-world CPR dynamics, enabling the study of how agents balance exploitation and conservation when faced with resources that differ in both productivity and management requirements.

\paragraph{Multi-agent reinforcement learning}

To investigate how domestication might emerge, we consider a small population of agents ${\cal P}=\{ ag^1,...,ag^N \}$ interacting sequentially with the environment $\cal E$ described above. At a given time step $t$ the environment is fully defined by its state $s_t\in S$. We use a Multi-Agent Reinforcement Learning (MARL) approach \cite{sutton2018reinforcement,hernandez2019survey,zhang2021multi,busoniu2008comprehensive, Littman1994} where each agent $ag_i$ is located on a given grid cell at time $t$ and can execute one of the following actions (Fig. 1d):
(1) move in one of the four directions to nearest grid points; (2) pick water or a seed and place it in its inventory, enabling the agent to then move with this item; (3) drop an item from its inventory on the current agent location; (4) harvest a plant and (5) perform a protection action within a $(3 \times 3)$ Moore neighbourhood. These 5 actions combined with their target item form a multi-discrete action space $A$ available to each agent (The complete set of actions available to agents at each time step—move, pick, drop, harvest, and protect—is described in SM I.C.). After executing $a^i_t\in A$, each agent $ag^i$ receives an immediate reward at time $t$: $r^i_t=R(s_t, a^i_t)\in\mathbb{R}$, i.e. positive rewards obtained when consuming a given plant and small negative rewards corresponding to action costs. The environment then transitions to the next state $s_{t+1}$ according to the state transition function $T: S\times A^N \rightarrow S$ specifying how the environment state is modified according to the environment dynamics and the actions executed by each agent. This process is repeated for the duration of an episode of 1024 time steps. At the end of an episode the environment is reset to initial conditions $s_0\in S$, where the positions of agents and plants are sampled probabilistically, and a new episode starts. Structuring the simulation as a sequence of episodes is required to train the agents using Reinforcement Learning, as described below. Each simulation run consists in a sequence of $1e6$ training episodes.



\paragraph{Individual reinforcement learning}

The objective of each agent $ag^i$ is to maximize its own cumulative reward during the duration of an episode, formalized as the discounted return (Fig.~\ref{fig:setup}a):
\begin{equation}
G^i = \sum_{t=0}^T \gamma^t R(s_{t}, a_{t}^i),
\end{equation}
where $T$ is the duration of an episode and $R:S \times A \rightarrow \mathbb{R}$ is the reward function. $\gamma \in [0,1]$ is the discount factor that balances the importance of immediate versus long-term rewards~\cite{sutton2018reinforcement}. Intuitively, $G^i$ quantifies the cumulative experience of agent $ag^i$, weighting earlier and later rewards according to $\gamma$. We specify long episodes of $T=1024$ steps corresponding to 40 full seasonal cycles, such that agents can potentially learn long-term eco-engineering strategies over multiple seasons.

This optimization problem is solved using reinforcement learning (RL) to train each agent action policy through trial and error. At each time step $t$, each agent $ag^i$ receives a partial observation of the environment state $obs^i(s_t)$ corresponding to its own local observation of plants, other agents and water in a $11\times 11$ neighbourhood (Fig.~\ref{fig:setup}-d), as well as the state of its inventory and the time of the season. In standard RL, the objective of an agent is to learn an action policy $\pi^i: O\rightarrow A$ mapping its observations $obs^i(s)\in O$ to actions $a\in A$, in order to maximize the cumulative return $G$.

In this work, we rely on a state-of-the-art RL method where each agent action policy is implemented as a Transformer neural network taking instead as input the whole sequence of \textit{(observations, actions, rewards)} tuples since the start of the episode and returning a probability distribution over actions~\cite{parisotto2020stabilizing, hamon:hal-04659863v1}, see MM for detail. The main advantage of this approach is to enable agents to integrate information through time and space over the current episode in order to best decide on the optimal next action.

Training is decentralized: each agent independently optimizes its own policy, relying solely on its individual interaction history without access to the observations, actions, or rewards of others. Therefore, as agents adapt simultaneously in a shared environment, the environment dynamics becomes non-stationary from the point of view of an individual agent. 
Each agent's policy training is performed using Proximal Policy Optimization (PPO)~\cite{schulman2017proximal}. Intuitively, PPO allows policy updates that improve performance but penalizes excessively large deviations, thereby stabilizing learning (see MM). 


 
\section{Results}


Using the MARL framework, we develop a progressive analysis of the emergence and stability of agriculture in learning populations. We first identify the conditions under which a small group of agents can discover agriculture through reinforcement learning. We then characterize the learning dynamics and the effective behavioral rules underlying this transition. We show that agriculture displays strong path dependence and becomes locked in once established. Extending to larger populations, we find that agricultural systems are vulnerable to cheaters but that learning dynamics can act as an endogenous firewall that protects cooperative regimes.

\begin{figure*}[t]
\begin{center}
    \includegraphics[width=0.95\linewidth]{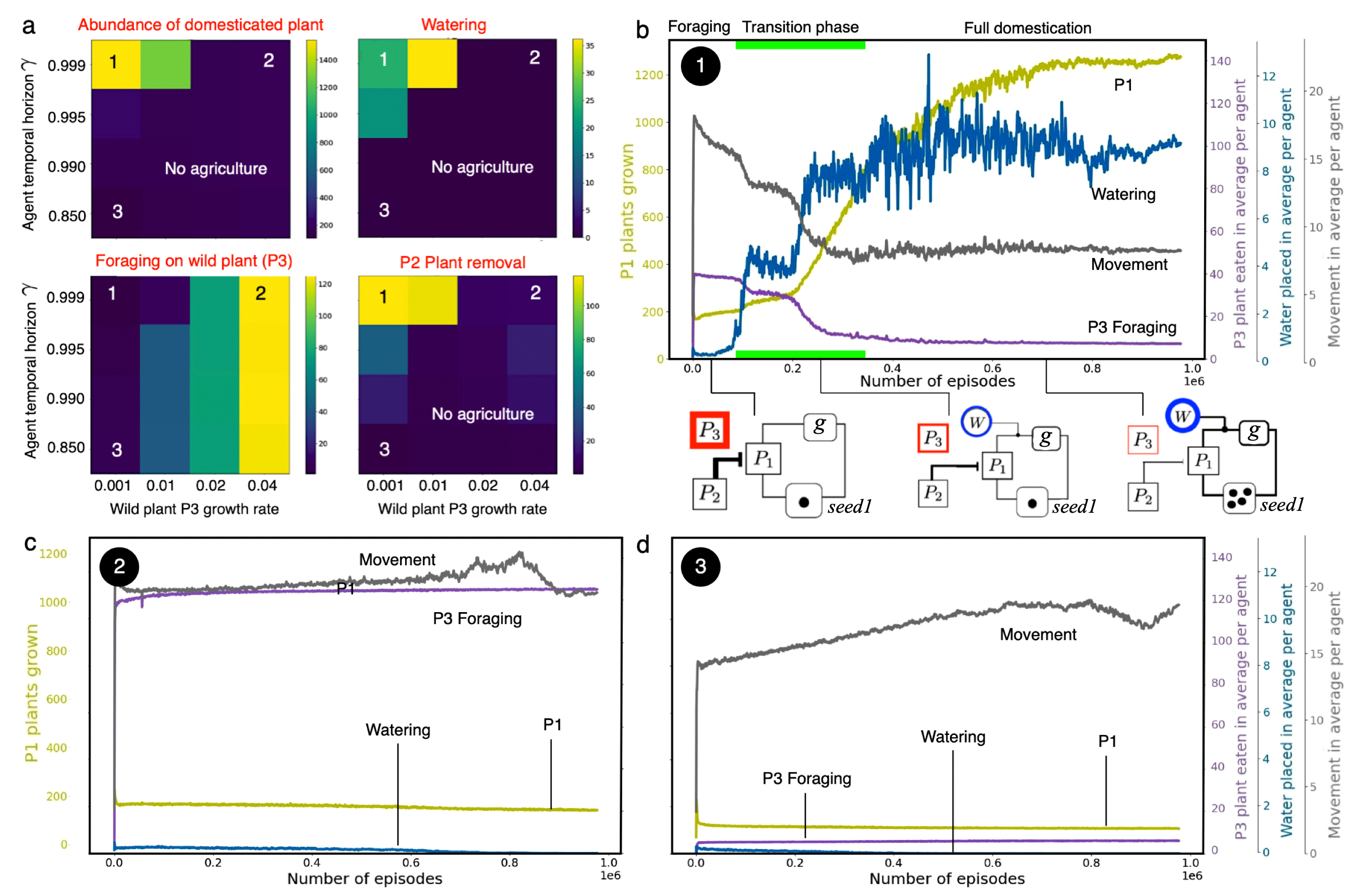}
\end{center}
\caption{{\bf Parameter-dependent emergence of agriculture}. In (a) we display four characteristic measures and their values for our MARL simulation model. Two key parameters have been used here: the rate at which the wild plant $P_3$ grows and the "cognitive" dimension of agents as captured by the discount rate $\gamma$, which defines the learning time horizon. In each parameter space we show the impact of each parameter combination (in this discretized $4 \times 4$ space) on different measures at the end of the simulation runs. All of them are consistent with a double requirement in order to evolve agriculture, namely a high $\gamma$ and low rates of $P_3$ growth. 
Additional behavioral metrics across the ($\gamma$, $P_3$ growth) parameter space are shown in Fig. S1 (SM III.A), including sustainability, protect action usage, episode reward. 
Three examples of the MARL dynamics are shown in (b-d), associated with the three combinations displayed in the parameter spaces (1, 2 and 3 in Fig.~\ref{fig:agri_main}.a). Each one shows how multiple measures evolve during agent's training (i.e. during the $1e6$ training episodes). In (b) an example of successful agriculture is shown, where we distinguish three phases: a foraging phase, a transition phase and a full domestication phase. Several relevant measures are displayed, including the total number of domesticated plants and the use of water for fertilization, both growing over time, and two declining indicators: agent movement and foraging. c) and d) show examples where agriculture did not emerge, either because $P_3$ growth rate was too high (c) or because $\gamma$ was to low (d). See SM, section I.D, for the precise definitions of all measures.
}
\label{fig:agri_main}
\end{figure*}

\subsection{Environmental and cognitive factors allowing the discovery of agriculture}

Two important components of agricultural success have been argued to be a reduction in foraging efficiency due to environmental stress (such as derived from persistent drought) and an evolved cognitive complexity that allows planning and tool development \cite{mithen1996prehistory}. To incorporate these two factors, we vary two key hyperparameters of the model. The first is environmental: the growth probability of the foraging resource ($P_3$). We hypothesize that agriculture is more likely to emerge when this probability is low, since cultivation, despite being costly, provides a clear cumulative advantage over foraging. The second is cognitive: the discount factor $\gamma$ that governs how agents value future rewards. A reasonable guess suggests that higher values promote agriculture discovery, as this strategy requires the ability to base decisions on long-term outcomes.

In Fig. 2a, the statistical results of our hyperparameter analysis are shown, using the parameter space defined by $\gamma$ against the probability of germination of the wild plant ($P_3$), $\eta_3$. Four different metrics have been used to determine the structure of this space, which reveals a domain of high $\gamma$ and low $\eta_3$ compatible with the emergence of agriculture. These are: the abundance of $P_1$, which measures the impact of domestication in population terms; the foraging on $P_3$, which weighs the relative importance of wild versus domesticated plants, and two metrics that capture the presence of ecological engineering, namely watering and the rate of removal of the strong competitor ($P_2$). Here, three case studies are marked, indicated as: (1) high-$\gamma$, low $\eta_3$, 
 (2) high-$\gamma$, high $\eta_3$ and (3) low-$\gamma$, low $\eta_3$, respectively.

An example of a successful outcome is shown in Fig.~2(b), where the 
time evolution of the MARL model is displayed. The system reaches 
a hunter--gatherer scenario during the first training episodes, exploiting $P_3$ (\emph{Foraging} phase). As time proceeds, a marked set of changes takes place within a \emph{Transition phase} (marked with an horizontal bar), 
as agents discover how to increase the abundance of $P_1$, which exhibits 
a steep growth over the course of the simulation through several 
synergistic processes.

A notable pattern is the reduction in agent movement as domestication progresses. This mirrors the behavior of human groups, whose reliance 
on wide-ranging foraging declines once agriculture makes sedentary 
life more advantageous. Consequently, 
the exploitation of $P_3$ decreases to very low levels, while the use 
of water to improve the soil and promote the germination of $P_1$ 
increases dramatically, leading to rapid agricultural expansion culminating at the \emph{Full domestication} phase.  

Below the time series panel (b), we summarize the changes that occur using a diagrammatic representation of the causal loops shown in Fig.~1(a). Here, we highlight the strength of interactions and dominant components using thick contours.
As indicated, while $P_3$ is initially exploited by foragers at $t=0$ and 
exerts strong pressure on $P_1$, this situation changes during the transition phase as agents increasingly manage watering, while foraging (and movement) markedly decline. In the fully domesticated regime (after $\sim 5e5$ episodes), watering is heavily used, agents have become sedentary, and a large yield of domesticated plants is present. This is in sharp contrast with the two other case studies within our parameter space, displayed in Fig.~\ref{fig:agri_main}(c) and (d) for which agriculture does not appear but for two very different reasons. In the case of panel c (corresponding to condition 2 in the parameter space) we are in a setting of fast growth of the foraging plan $P_3$, thus lowering the pressure for the discovery of the potential power of $P_1$ being cultivated. This is clearly the case as in both situations 1 and 2 the agent temporal horizon is the same ($\gamma=0.999$). In this case, we observe in the plot the constant high consumption of the foraging plant, also the high rate of agent movement, and very reduced watering of plants. On the other hand, Fig.~\ref{fig:agri_main}(d) depicts a situation in which the foraging plant $P_3$ is not abundant, but agriculture is not discovered as it requires a longer temporal horizon for decision making (controlled by parameter $\gamma$). In this situation, agents also move at a high rate to collect the very scarce foraging resource as seen in the plot.

\begin{figure*}[t]
\begin{center}
    \includegraphics[width=0.99\linewidth]{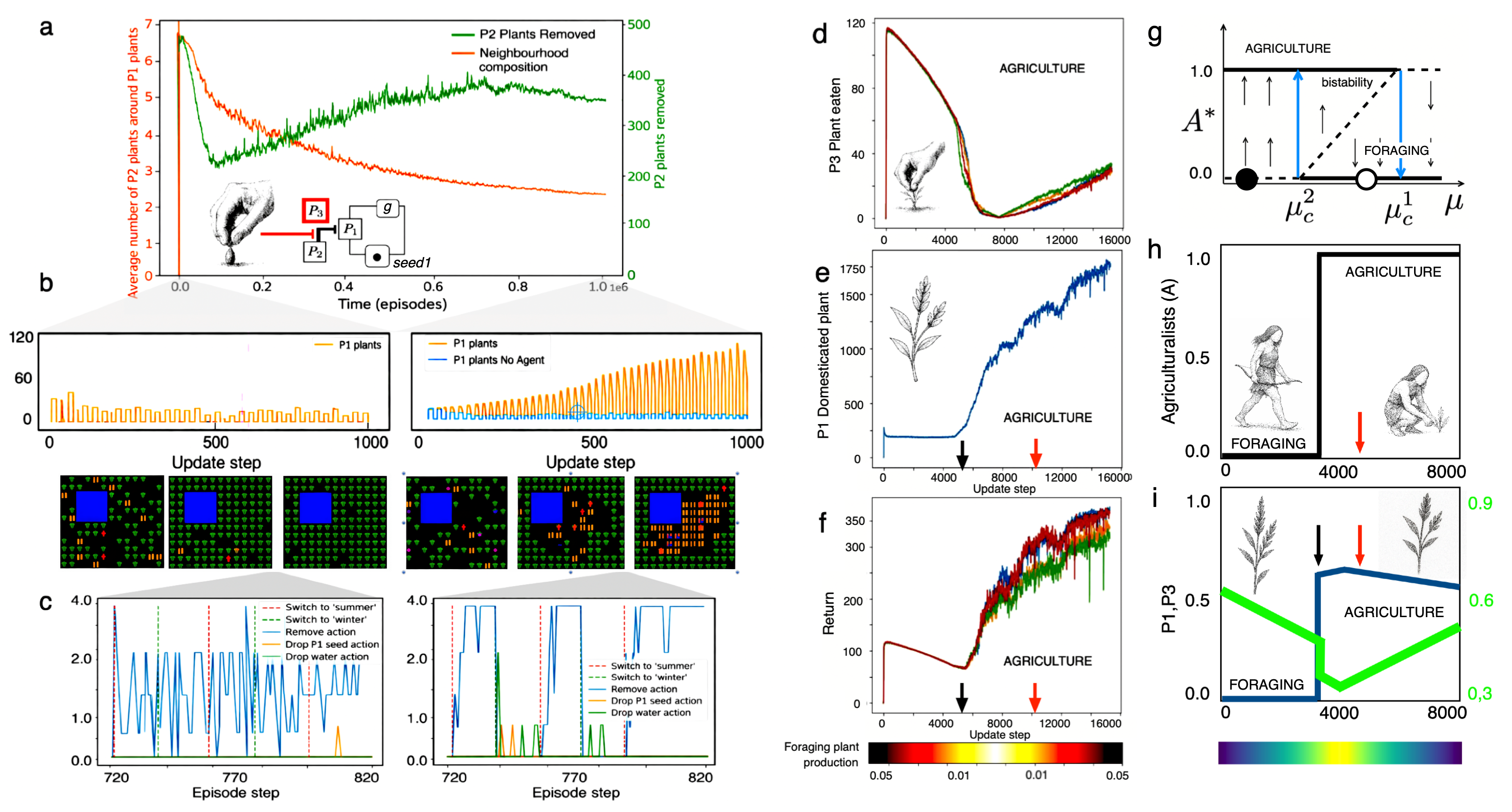}
\end{center}
\caption{{\bf Scales, transitions and lock-in states.} 
(a) Over $10^6$ episodes, $N=4$ agents learn an effective strategy of ecological engineering by removing the competitive $P_2$ weed, reducing its density around the rewarding $P_1$ plant. 
(b) Within-episode dynamics of $P_1$ for untrained agents (left) and fully trained agents (right). The right plot also shows P1 dynamics when no agent is present. Seasonal oscillations are visible, but only trained agents strongly amplify $P_1$ by actively cultivating it near the water source (snapshots below showing the simulation environment, with water source in blue, agents in red, $P_1$ in yellow, $P_2$ in green and $P_3$ in purple). 
(c) Action distributions within a few seasons show that untrained agents (left) behave similarly across seasons, whereas trained agents adopt a clear seasonal strategy: removing $P_2$ in summer and dispersing water and $P_1$ seeds in winter. 
(d, e, f) Evolution of measures in a condition where we progressively vary the availability of the $P_3$ wild plant from high to low to high values (as indicated by the color bar at the bottom). Agents learn a foraging strategy when $P_3$ is initially abundant, switch to agriculture when it becomes scarce (black arrow), but do not return to foraging when abundance is back (red arrow). 
(g, h, i) A mean-field (MF) mathematical model was derived (see Materials and Methods and SM) from the MARL model results, confirming the presence of hysteresis, as indicated in (g), when tuning $\mu$ from a given value to a lower one and back ($\circ\rightarrow \bullet \rightarrow \circ$ in (g)). In this framework, agriculture is an all-or-none phenomenon (i.e.\ $A^*=0$ and $A^*=1$ are the two stable equilibrium states). 
In (h) and (i) we show the mean field model dynamics (see Materials and Methods) for the fraction of agriculturalists $A$, and the abundances of $P_1$ and $P_3$, under a slow triangular forcing of the environmental parameter $\mu$, which controls the spontaneous regeneration rate of the wild plant. Specifically, $\mu$ is decreased linearly from $\mu_{\max}=1.2$ to $\mu_{\min}=0.8$ over $T_{\downarrow}=4000$ time units and then increased back to $\mu_{\max}$ over $T_{\uparrow}=4000$, with all other parameters fixed ($\eta_F=1.0$, $\eta_A=1.4$, $\kappa=1.0$, $\Delta f_0=-0.6$, $v=0.35$, $\chi=1.0$, $u=0.6$, $\omega=1.0$, $c_u=0.9$, $c_v=2.0$, $c_s=1.0$). The system is initialized near the pure foraging equilibrium. As $\mu$ decreases, the system undergoes a discontinuous transition from a foraging state ($A\approx 0$) to an agricultural state ($A\approx 1$), as indicated by the black arrow. When $\mu$ is subsequently increased, no reverse transition occurs when the previous conditions are met again (red arrow): a hysteresis loop is present. In contrast to the continuous amplification of $P_1$ observed in the MARL simulations (e), the MF model  predicts a sharp regime shift followed by a decay in $P_1$.} 
\label{fig:scales}
\end{figure*}

To quantify how learning reshapes the coupled dynamics of agents and vegetation, we analyze the system across multiple temporal scales, from the full training horizon down to individual seasonal cycles. As summarized in Fig.~\ref{fig:scales}(a), over long time scales, agents progressively discover an effective form of ecological engineering: by selectively removing the competitive $P_2$ weed, they suppress its local density around the rewarding $P_1$ plant. To measure the effect of selective removal of $P_2$ plants, we averaged over the grid using the average number $\langle n_1(t) \rangle $ of $P_2$ plants surrounding a domesticated $P_1$ one (\emph{Neighbourhood composition} in Fig~\ref{fig:scales}a). If we indicate by $\bf r$ the location of site in $\Omega$, and $\Gamma({\bf r})$ is the set of nearest neighbours, this gives us: 
\begin{equation}
\langle n_1(t) \rangle = \left < {1 \over N_1(t)} \sum_{{\bf r}\in \Omega} \delta_{S({\bf r}),P_1} \left ( \sum_{{\bf k} \in \Gamma(S({\bf r})} \delta_{S({\bf k}),P_2}\right ) \right >
\end{equation}
where $\delta_{ij}=1$ when $i=j$ and zero otherwise. Low values of $\langle n_1(t) \rangle $ indicates that agents mitigate plant competition to enhance resource access. This emergent strategy is reflected in the steady increase of $P_2$ removal events and the progressive spatial clearing that forms around $P_1$. 

On the scale of single episodes (Fig~\ref{fig:scales}b), untrained agents are essentially indistinguishable from those of an agent-free environment, with $P_1$ following merely passive seasonal oscillations. In contrast, fully trained agents can strongly amplify $P_1$ abundance within an episode by actively cultivating it near the water source. Finally, zooming into a few seasonal cycles reveals a profound reorganization of behavior: while untrained agents act in an unstructured and seasonally uncorrelated manner, trained agents develop a clear temporal division of labor, preferentially removing $P_2$ during summer and dispersing water and $P_1$ seeds during winter. This structured seasonal policy underlies the long-term ecological control achieved by the agents.

According to the assumptions of our model, the discovery of agriculture is favored by two important environmental and cognitive factors. First, agriculture is hardly discovered in environments with an abundant access to foraging resources, even though agriculture could still yield higher rewards if it were discovered. Our interpretation is that, in such conditions, a foraging strategy constitutes a strong local optima from which it is then hard to escape. Second, the discovery of agriculture requires a cognitive ability to make decisions based on long term predictions of future outcomes. With low discount factors, agents cannot learn how to eco-engineer their environment in the present, which implies a high immediate cost, in order to collect more rewards in the future.

\subsection{Domestication as a lock-in state}

In our previous examples, we fixed the key parameters that define the environment and allowed our MARL system to discover, when feasible, the paths to domestication. In this context, the transition to agriculture might have been favored by the loss of some foraging resource \cite{belfer2000early}, which in our model would involve a decrease in the presence of the foraging plant $P_3$. On the other hand, we can also ask whether our artificial farmers might transition back to foraging once the foraging plant returns to previous levels. Is that the case? Evidence suggests that once agriculture is established, the system becomes trapped in that state, both in human \cite{gowdy2016economic} and insect \cite{aanen2006social} societies. This is illustrated in Fig.~\ref{fig:scales}d-f, where we gradually changed the foraging plant production over the training episodes (as indicated in the color bar below). At a given value of the wild plant growth rate (black arrow), the MARL model shows that agriculture emerges as a solution to the decay of foraging resources. A marked increase in $P_1$ is observed in (e) along with a rapid growth in (f). These trends are persistent, even when the decay trend in $P_3$ is reversed: once we cross the previous transition point (red arrow) there is no coming back. 

These results indicate the presence of hysteresis. To examine its origin, we 
consider a reduced mathematical model guided by the RL results. As described in 
\textit{Materials and Methods}, the model incorporates the observed cost--benefit 
trade-offs but assumes fixed parameters (i.e., no learning). The dynamical 
structure can therefore be analyzed using bifurcation theory 
\cite{strogatz2024nonlinear,sole2011phase}. Let $A$ and $F$ denote the fractions of agriculturalists and foragers, with 
$A+F=1$. Introducing the parameter $\mu$, which controls the spontaneous growth 
of $P_3$ (all other parameters fixed), we analyze a dynamical sequence ($\circ\rightarrow \bullet \rightarrow \circ$, as indicated in Fig.~\ref{fig:scales}g) where the system starts from the 
foraging state $\circ:=(F=1,A=0)$. Decreasing $\mu$ produces no qualitative change until 
a bifurcation occurs at $\mu=\mu_c^2$, where the system abruptly transitions to 
the agricultural state $\bullet:=(A=1,F=0)$ (Fig.~\ref{fig:scales}g--h). Reversing the change in $\mu$ does not immediately restore the foraging state: 
the system remains trapped in the agricultural attractor even when $P_3$ returns 
to previous levels (red arrows). Only when $\mu$ exceeds a second threshold 
$\mu=\mu_c^1$ (Fig.~\ref{fig:scales}g) does the system recover the foraging state, generating 
the observed hysteresis. The absence of learning leads to a markedly different resource dynamics 
(Fig.~\ref{fig:scales}i). Once the agricultural attractor is reached, $P_1$ increases sharply 
but then slowly decreases as $\mu$ grows (and more foraging plants become available). In contrast, the RL dynamics (Fig.~\ref{fig:scales}e) actively reshapes the system: learning maintains and expands $P_1$ 
despite the increasing presence of $P_3$, stabilizing the agricultural state.


\subsection{Social learning as a firewall against cheaters}

Domestication emerges in our MARL system even with a small number of agents, which we kept fixed in their numbers in our previous simulations. This result is notable because it shows that agricultural behavior can arise 
from individual learning alone, without requiring large populations or 
pre-existing sociocultural exchanges. However, the cooperative loop underlying 
the domestication process relies on shared resources and is therefore vulnerable to the 
appearance of cheaters. This effect is illustrated in Fig.~4a, where the same quantities analyzed in 
Fig.~2a are explored as a function of the number of agents. The results show 
that agriculture emerges only when the population size remains below a 
critical threshold $N \leq N_c \sim 5$. For larger populations, the 
increasing presence of cheaters disrupts cooperative dynamics and 
prevents the stabilization of agricultural strategies. Some improvement can be obtained when changing the initial conditions at the start of episodes, if the previous positions of the agents are kept. In that case, the threshold is pushed to $N_c \sim 10$ but beyond this cheaters also emerge (see SM). Is there a way to keep a stable cooperation in larger populations?

\begin{figure*}[t]
\begin{center}
    \includegraphics[width=0.9\linewidth]{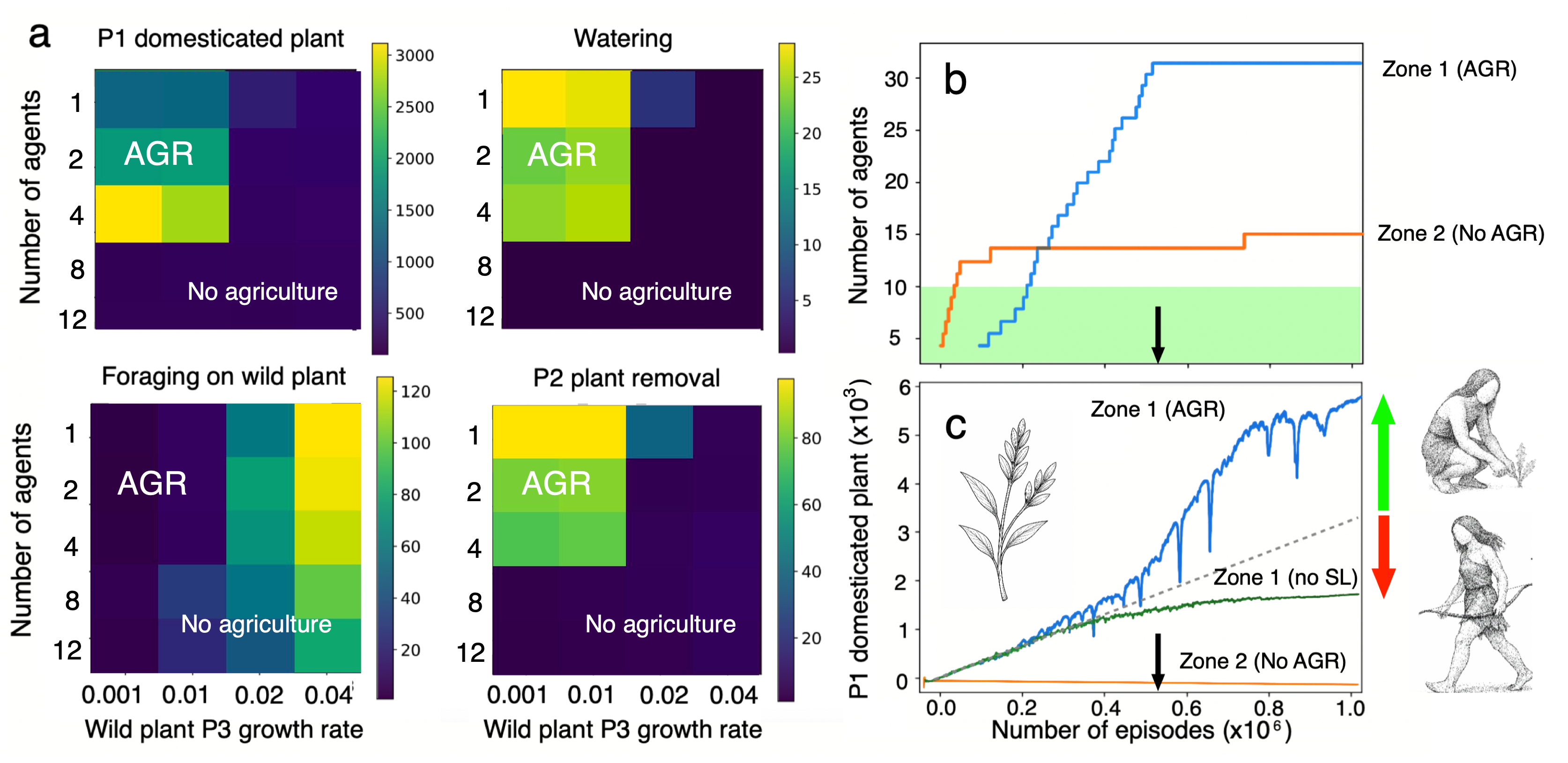}
\end{center}
\caption{{\bf Population dynamics growth under social cloning}. (a) Heat maps showing the outcome of simulations as a function of the wild 
plant growth rate $P_3$ and the number of agents. The panels report the 
abundance of the domesticated plant $P_1$, watering activity, foraging on the 
wild plant $P_3$, and removal of the competing plant $P_2$ (the three last measures being reported as the maximum over agents). Agriculture (AGR) 
emerges only within a restricted region of parameter space corresponding to 
small populations and sufficiently low $P_3$ growth rates. For larger 
populations ($N \gtrsim N_c \sim 5$), agriculture fails to appear and the 
system remains in a foraging-dominated regime (``No agriculture''). This 
transition reflects the increasing impact of cheaters: agents that exploit the 
benefits of cultivated resources without contributing to costly agricultural 
actions, thereby suppressing the cooperative behaviors required for plant 
domestication. (b) Population dynamics when cloning (policy inheritance) is allowed. In the 
agricultural region (Zone 1, Fig.2a-d) the number of agents under SL grows 
over time (with a maximum $N_{\max}=32$ allowed in our simulations), whereas it remains 
low $N=15$ and flat when agriculture does not emerge. (c) Time series of the domesticated plant $P_1$. In the agricultural scenario, $P_1$ exhibits nonlinear growth that exceeds the linear baseline prediction 
(dashed line), reflecting positive feedback between cultivation actions and 
plant growth, even after population saturation occurs (black arrow). By contrast, when cloning is suppressed (no SL), the abundance of $P_1$ increases only sublinearly and eventually 
saturates. Allowing cloning therefore overcomes the cheater-limited regime 
observed in (a) by enabling successful agricultural strategies to spread in 
the population, leading to sustained growth.
}
\label{fig:population}
\end{figure*}

To answer this question, we need to incorporate social learning (SL), as supported by 
archeological evidence, suggesting that early cultivation practices spread mainly through cultural transmission 
\cite{feldman1996individual,mesoudi2006towards,mesoudi2016evolution,demps2024mind}. Domestication and early 
farming involved cumulative, multi-step practices difficult to propagate through individual trial-and-error alone. In a multi-agent reinforcement learning (MARL) framework, social learning allows successful behaviors 
discovered by some agents to spread through the population, reproducing the diffusion and cumulative cultural 
dynamics observed through the Neolithic transition 
\cite{sterelny2011hominins,jerardino2014cultural,fort2023neolithic}. 

To represent this process, we introduce a reproduction mechanism in which new agents inherit the behavioral policy of successful individuals—a minimal surrogate for social learning within MARL that allows effective strategies to persist and spread throughout the population~\cite{leibo2019malthusian}. Crucially, this mechanism is strategy-agnostic: cheaters are cloned with equal opportunities than agriculturalists, thus, short-term exploitative behavior can spread just as readily as cooperative farming practices. This captures a key feature of human subsistence systems, where complex practices such as agriculture depend on the cumulative transmission of knowledge across generations.

Simulations start with an initial population size $N_0$. In the simulation, agents reproduce whenever their cumulative reward exceeds an upper threshold 
and die when it falls below a lower threshold. When reproduction occurs, a new agent is introduced at the next episode with the same action policy as its parent, effectively transmitting the learned 
strategy. The population is capped at a maximum size $N_{\max}$ ($=32$ in Fig.4b due to computational limitations), beyond which reproduction is no longer possible. All other hyperparameters remain identical 
to those used in the previous section (fixed environment parameters, training hyperparameters and neural network architecture details are reported in Tables S.I and S.II respectively of SM).

Let the population at episode $t$ consist of $N_t$ agents indexed by 
$i=1,\dots,N_t$. Each agent is characterized by a policy $\pi_i$ and a 
cumulative reward $R_i(t)$. Reproduction and death are controlled by two 
thresholds $R^{+}$ and $R^{-}$. If $R_i(t) \ge R^{+}$ and $N_t < N_{\max}$,
agent $i$ reproduces and a new agent is introduced at the next episode with 
the inherited policy
\[
\pi_{N_t+1}(t+1) = \pi_i(t).
\]
Conversely, if $R_i(t) \le R^{-}$, the agent is removed from the population.
Denoting by $\mathcal{B}(t)$ the set of agents satisfying the reproduction 
condition and by $\mathcal{D}(t)$ those satisfying the death condition, the 
population size evolves as $N_{t+1} = N_t + |\mathcal{B}(t)| - |\mathcal{D}(t)|$, provided that 
$N_{t+1} \le N_{\max}$. This cloning mechanism allows policies associated with high reward to 
propagate. Figure 4(b-c) illustrates how cheating is overcome once cloning 
(policy inheritance) is allowed. In this case, successful policies are 
replicated through reproduction, so that agents implementing effective 
agricultural strategies generate descendants that inherit the same behavioral 
rules. As a consequence, episodes in which cultivation behaviors become 
dominant lead to a gradual increase in population size, whereas in the 
non-agricultural regime the population remains approximately constant. The dynamics of the domesticated plant $P_1$ highlights the importance of this feedback. When agriculture emerges, the abundance of $P_1$ displays a strongly 
nonlinear increase that clearly exceeds the linear baseline prediction 
(dashed line), reflecting the positive feedback between cultivation effort, resource availability, and population growth. By contrast, when social learning is not at work (no SL, Fig.4c), the growth of $P_1$ remains sublinear and eventually saturates. The propagation of agricultural practices through cloning 
allows the cumulative retention of techniques such as 
watering, selective harvesting, and the removal of competing plants. In summary, population size under SL is a firewall against the emergence of cheaters.

Additional results from the social learning experiments, including the effect of the spontaneous growth rate of the foraging plant (P3) on population survival, are reported in SM III.C (Fig. S3).

 
\section{Discussion}


The emergence of agriculture represents a major evolutionary transition shaped
by environmental constraints. Archeological evidence indicates that
domestication rarely occurred in resource-rich settings, but instead in variable
environments where foraging alone could not reliably sustain populations \cite{flannery1973origins,smith2001low,richerson2001cultural,cohen1977population}. Although previous 
agent-based models have been used to study cultural innovations \cite{epstein1996growing,epstein2000understanding,lansing2003complex,lansing2012perfect,demps2024mind}, our results provide a novel mechanistic account of this transition by showing how
feedbacks between resource availability, learning dynamics, and environmental
modification generate tipping points between foraging and farming regimes.

Our model identifies four key ingredients that underlie the emergence of
domestication. First, agriculture requires planning at the individual
level: agents must value delayed rewards to invest in costly actions such as
watering or seed dispersal. We model this using the discount factor $\gamma$, which allows us to
quantify the present value of future rewards. Second, the transition is inherently \emph{social}:
cultivation relies on shared resources and is therefore vulnerable to the
emergence of cheaters, which destabilize cooperative investment. Third, and
critically, this limitation can be overcome by introducing \emph{social
learning}. As shown in Fig.~4, agriculture fails above a critical population
size due to cheater invasion, but is restored when policy inheritance allows
successful strategies to propagate. In this regime, social learning acts as a
\emph{firewall}, stabilizing cooperation, and enabling sustained population
growth. Fourth, once established, agriculture exhibits an effectively 
irreversible or lock-in character: systems that transition from
resource-poor environments to cultivation remain trapped in a
high-productivity agricultural state due to strong positive feedbacks between
population size, environmental modification, and resource abundance. The
resulting dynamics exhibit nonlinear amplification of the domesticated plant,
reflecting the coupled effects of cultural transmission, resource management,
and ecological feedbacks. This supports the importance of the social pathway, along with the 
agent-environment feedback, towards complex AI \cite{duenez2023social,hertz2025beyond}.

Beyond agriculture, this framework provides a general perspective on the emergence of \emph{cultural innovations}. 
Complex behaviors such as cultivation arise without explicit instruction, emerging instead from the coupling between 
reinforcement learning and ecological dynamics, and are subsequently stabilized through social transmission. This 
minimal form of cumulative culture demonstrates that collective learning systems can generate, retain, and propagate 
innovations even in the absence of sophisticated cognition or explicit communication. More broadly, our results 
highlight the potential of multi-agent artificial intelligence as a powerful approach to investigate how collective solutions can emerge in a context where agents can modify their environments to reduce the uncertainty of their worlds. Although our models are simple constructs, they offer a proof of concept for the potential of AI systems to address timely challenges regarding human-environment interactions \cite{sole2022ecological}. 

Several limitations should be noted. The model does not incorporate \emph{resource storage}, which is likely critical 
for buffering seasonal variability and stabilizing agricultural strategies. In addition, plant populations 
lack genetic variability, precluding co-evolutionary feedbacks associated with domestication, 
such as selection for increased yield 
or reduced dispersal. Social learning is also represented in a minimal form through cloning, capturing transmission but not richer processes such as teaching or the recombination of innovations. Despite these 
limitations, the framework offers a tractable platform for studying broader cognitive transitions 
\cite{szathmary2015toward,sole2016synthetic,ginsburg2021evolutionary,sole2026cognition}. In particular, it opens the 
possibility of testing whether similar mechanisms underlie the emergence of ultrasociality in ants and termites 
\cite{gowdy2013ultrasocial,gowdy2016economic,schultz2022convergent}, where large-scale cooperation arises from the interplay among ecological constraints, division of labor, and collective information processing (see SM, section
V). In a related context, our model approach can also be extended considering interactions between competing communities to test hypotheses regarding the role played by conflict in the evolution of ultrasociety in terms of norms and institutions \cite{turchin2007war,bowles2011cooperative,Turchin2016Ultrasociety}.


\section{Material and methods}


\paragraph{Environmental dynamics}

The general dynamics of the simulation environment are described in Section~\ref{sec:methods}. Here we provide the formal details required for reproducibility and refer to \textit{Supplementary Material} (SM) the definition of all parameter values (sections I.A and I.B in SM). The simulated environment consists of a $30\times30$ discrete grid, corresponding to 900 cells arranged on a two-dimensional spatial landscape. Each cell is identified by coordinates $(x,y)\in[1,30]^2$ and stores the following state variables: (a) $ag^{x,y}$, the number of agents present in the cell; (b) $seeds_1^{x,y}$, the number of $P_1$ plant seeds; (c) $growth_1^{x,y}$, the growth state of plant $P_1$ in the cell, ranging from 0 (no plant present) to 1 (fully grown); (d) $seeds_2^{x,y}$ and $growth_2^{x,y}$, defined analogously for plant $P_2$; (e) $water^{x,y}$, a Boolean variable that indicates whether the cell contains a water source; (f) $watered^{x,y}$, a Boolean variable indicating whether the cell has been watered by an agent (see SM for more detail); (g) $W^{x,y}_{time}$ the amount of time steps $watered^{x,y}$ has been to True during the current season; and (h) $P_3^{x,y}$, a Boolean variable indicating the presence of plant $P_3$ in the cell. The temporal dynamics of the grid is driven by seasonal cycles. As mentioned in Section~\ref{sec:methods}, a simulation run consists in a sequence of $10^6$ episodes of 1024 time steps each. An episode consists of 25 seasonal cycles. Each seasonal cycle consists of a summer season followed by a winter season. Each season lasts 20 time steps. At the start of each episode, the grid is initialized with random positions of agents and seeds and a fixed $5\times 5$ water source in the middle of map (see SM for detail). Each episode starts with the summer season. At the beginning of each summer, seeds can germinate, following a saturating form:
\[
germ_1^{x,y} = \frac{\alpha_1 seeds_1^{x,y}}{1+\beta_{21} seeds_2^{x,y}}, 
\;\;\;
germ_2^{x,y} = \frac{\alpha_2 seeds_2^{x,y}}{1+\beta_{12} seeds_1^{x,y}},
\]
where $\alpha_i$ are baseline germination rates and $\beta_{ij}$ measures competitive inhibition between seed types. The probability that a cell $(x,y)$ becomes occupied is limited by space availability. We therefore define:
\begin{equation}
z_1^{x,y} = clip({germ_1^{x,y}},0,1), \;\;\;
z_2^{x,y} = clip({germ_2^{x,y}},0,1), \;\;\;
\end{equation}
\begin{equation}
z_{\emptyset}^{x,y} = 1-clip(z_1+z_2,0,1).
\end{equation}
and 
\begin{equation}
p_1^{x,y} = \frac{(z_1^{x,y})^{temp}}{Z^{x,y} }, \;\;\;
p_2^{x,y} = \frac{(z_2^{x,y})^{temp}}{Z^{x,y} }, \;\;\;
p_{\emptyset}^{x,y} = \frac{(z_{\emptyset}^{x,y})^{temp}}{Z^{x,y} }.
\end{equation}
where $temp$ is an inverse temperature parameter and $Z^{x,y}=(z_1^{x,y})^{temp}+(z_2^{x,y})^{temp}+(z_{\emptyset}^{x,y})^{temp}$ so that $p_1+p_2+p_{\emptyset}=1$, forming a discrete probability distribution. At the beginning of a summer season, one of these three outcomes is sampled independently for each cell $(x,y)$, deciding if the cell will germinate either the plant $P_1$, $P_2$ or none of them. During the last 10 time steps of summer, plants disperse seeds into their $3\times3$ neighborhood. For each cell $(x,y)$, let's note $n_i^{x,y}$ the number of neighboring cells occupied by $P_i$, i.e.:
\begin{equation}
    P_i^{x,y} = \mathbf{1}_{growth_i^{x,y} > 0} \;\;\; \;\;\; 
\end{equation}
\begin{equation}
n_i^{x,y} = \sum_{(\hat{x}, \hat{y})\in \mathcal{N}(x,y)}(1 + \mathbf{1}_{i=1} watered^{\hat{x},\hat{y}})P_i^{\hat{x}, \hat{y}},
\end{equation}
where $\mathbf{1}_{cond}$ is the indicator function equal to one if $cond$ is true, 0 otherwise ; and $\mathcal{N}(x,y)$ is the set of cells in the Moore neighbourhood of cell $(x,y)$, including itself.
The probability that a plant of type $P_i$ produces a seed in a focal cell $(x,y)$ is $\rho_i^{i,j} = \gamma_i (n_i^{x,y}+P_i^{x,y})$, where $\gamma_i$ is the per-neighbor dispersal rate. Thus, $P_1$ plants that are on a watered cell have higher probability to disperse seeds compared to those that are not.
At the beginning of winter, all existing plants die and only 
seeds remain. Seeds that do not germinate may disappear before the next summer with a 
fixed survival probability $d_{seed}$. In addition, each cell can spontaneously generate a wild 
plant $P_3$, which replaces $P_1$ or $P_2$ if present. During summer, plant nutritional 
value increases smoothly with maturity. The harvest reward of a $P_1$ on cell $(x,y)$ is modeled using a 
standard logistic function,
\[
R_{P_1} = 
\frac{1}{1+\exp\!\left[-\xi\left(growth_1^{x,y}-1\right)\right]}
\left(1+W_{\text{time}}^{x,y}\right).
\]
Harvesting $P_2$ does not provide any reward, while harvesting $P_3$ provides a fixed reward $R_{P_3}=\mathbf{1}_{P_3^{x,y}}$ (i.e. 1 if when a $P_3$ is present). Harvesting a plant has a fixed cost $c_{harvest}$ subtracted to the reward. Thus, the total reward received by an agent at a time step when it is on cell $(x,y)$ is given by $R= \left( R_{P_1} + R_{P_3}- c_{harvest} \right)\mathbf{1}_{a_{harvest}}$, where $\mathbf{1}_{a_{harvest}}$ is 1 if the agent use the harvest action and 0 otherwise.

\paragraph{Transformer-based action policy}

Each agent $ag_i$ in the simulation is equipped with an \textit{action policy}:
\begin{equation}
    \pi_\theta^i : (O\times A \times \mathbb{R})^{T_{mem}} \rightarrow \Delta(A), \;\;\;\;
h_t^i \mapsto \pi_\theta^i(h_t^i),
\end{equation}
where $\theta$ are learnable parameters, $O$ is the observation space, $A$ is the action space, $\mathbb{R}$ is the space of scalar rewards and $\Delta(A)$ is the space of probability distributions over $A$ (see Section~\ref{sec:methods} for the definition of $O$ and $A$). Therefore, an action policy receives as input the history $h_t^i$ of \textit{(observation, action, reward)} tuples over the last $T_{mem}$ steps and returns a probability distribution over the available actions. At each time step $t$ of an episode, each agent $ag_i$ queries its own action policy $\pi_\theta^i$ with its current history $h_t^i$ to sample its next action $a\in A$, padding the missing inputs in $h_t^i$ with zeros when $t<T_{mem}$. The action policy of each agent is implemented as a Transformer neural network, where $\theta$ are the learnable parameters. At the start of a simulation run, all parameters are randomly sampled. Then they are trained from the data collected by each agent as explained below. The neural network architecture follows \cite{parisotto2020stabilizing}, see SM for hyperparameters. 

\paragraph{Training with Proximal Policy Optimization}

We use Proximal Policy Optimization (PPO) \cite{schulman2017proximal} to train the action policy $\pi_\theta^i$ of each agent. PPO is a policy gradient Reinforcement Learning algorithm, which updates the policy parameters $\theta$ in the direction of an estimated expected return gradient. Standard policy gradients may suffer from instability when policy updates are too large; PPO mitigates this by introducing a clipped surrogate objective:  
\begin{equation}
L^{CLIP}(\theta) = \hat{\mathbb{E}}_t \left[ 
\min \big( r_t(\theta)\hat{A}_t,\,
\text{clip}(r_t(\theta), 1-\epsilon, 1+\epsilon)\hat{A}_t \big) 
\right],
\end{equation}
where $r_t(\theta) = \pi_\theta(a_t|h_t)/\pi_{\theta_{\text{old}}}(a_t|h_t)$ is the ratio of new to old policy probabilities, $\hat{A}_t$ is an estimator of the advantage function, and $\epsilon$ is a small hyperparameter. 
To promote persistent exploration, the PPO loss is enhanced with an entropy bonus, which encourages the policy distribution to remain sufficiently stochastic. This prevents premature convergence to suboptimal strategies. The strength of this regularization is controlled by the parameter $\lambda_{\text{entr}}= 0.036$ as the default value if not stated otherwise. We used the implementation of \cite{hamon:hal-04659863v1} to efficiently train agent policies using GPU acceleration. The role of this exploration bonus and its influence on the discovery of agriculture is examined in SM III.D (Fig. S4).

\paragraph{Mean field model}

Simulations of the MARL model motivate a low-dimensional deterministic approximation that captures the key ecological and strategic tradeoffs without explicitly modeling learning. The formalization of the mean field model below uses different notation compared to the MARL model above. We track the fraction of agriculturalists $A(t)\in[0,1]$ (foragers $F=1-A$) and plant cover fractions $P_i(t)\in[0,1]$ satisfying $P_1+P_2+P_3=1$, where $P_1$ denotes domesticate, $P_2$ weed and $P_3$ wild foraging plants. Using the reduced state $(A,P_1,P_3)$ we set $P_2=1-P_1-P_3$.

The total encounter rate with wild plants is
\[
H(A)=\eta_F(1-A)+\eta_A A
=\eta_F+(\eta_A-\eta_F)A,
\]
where $\eta_F$ and $\eta_A$ are the encounter rates of foragers and farmers. Wild plant turnover follows
\[
\frac{dP_3}{dt}=\mu(1-P_3)-\kappa H(A)P_3,
\]
where $\mu$ is the turnover of the background vegetation and $\kappa$ measures the intensity of harvest per encounter.

Competition between domesticate and weed is biased by agriculture through facilitation,
\[
\Delta f(A)=\Delta f_0+ v\chi A,
\qquad
\Delta f_0:=f_1^0-f_2^0,
\]
where $f_1^0$ and $f_2^0$ are baseline fitness (intrinsic growth advantages) of domesticate and weed in the absence of cultivation, $\Delta f_0$ their baseline competitive difference, $v$ the intensity of niche construction, and $\chi$ the strength with which agriculture amplifies domesticate advantage. The $P_1$ dynamics read
\[
\frac{dP_1}{dt}
= P_1P_2\,\Delta f(A)
+ u\omega A\,P_2
- \mu P_1
+ \kappa H(A)P_3 \frac{P_1}{1-P_3},
\]
where $u$ is the effort devoted to the removal of weeds, $\omega$ its efficiency, and the last term represents the replacement of $P_2$  by $P_1$. The $A$ population follows a replicator equation,
\[
\frac{dA}{dt}=A(1-A)(\pi_A-\pi_F),
\]
with payoffs
$
\pi_A = f_1^0 P_1 + b_3 P_3 - c_u u\,P_2 - c_v v,$$
\pi_F = f_2^0 P_1 + b_3 P_3 - c_s s(P_3),$
Here $b_3$ is the nutritional benefit from wild plants shared by both strategies; $c_u$ and $c_v$ are costs of effort $u$ and facilitation $v$; $c_s$ scales foraging search costs; and $s(P_3)$ is a decreasing function of wild plant availability capturing increased search effort under scarcity. The system preserves $A,P_1,P_3\in[0,1]$ and enforces $P_1+P_2+P_3=1$ by construction.


\section*{Acknowledgments}

RS thanks the members of the Complex Systems Lab for their valuable discussions and the support by the AGAUR 2021 SGR 0075 grant, AEI-PID2023-152129NB-I00 grant and the Santa Fe Institute. MSF is supported by project ACISUD PID2022-136787NB-I00 and DEEPEEG PID2024-156859NA-100. CMF was supported by the Agence Nationale de la Recherche (ANR, grant ANR-20-CE23-0006, ECOCURL project). GH received funding from the UBGRS-Mob mobility grant (ANR, reference ANR-20-SFRI-0001). This work also benefited from access to the HPC resources of IDRIS under the allocation A0091011996 made by GENCI.

\bibliographystyle{unsrt} 
\bibliography{references}

\end{document}